\documentclass[aps,prl,twocolumn,superscriptaddress]{revtex4}

\usepackage{graphicx}
\usepackage{color}

\begin{document}


\title{Extreme vortex pinning in the non-centrosymmetric  superconductor CePt$_{3}$Si }



\author{C.F. Miclea}
\email[]{miclea@cpfs.mpg.de}
\affiliation{Max-Planck-Institute for Chemical Physics of Solids, Dresden, Germany}

\author{A.C. Mota}
\affiliation{Solid State Lab., ETH-Zurich, Zurich, Switzerland}

\author{M. Nicklas}
\affiliation{Max-Planck-Institute for Chemical Physics of Solids, Dresden, Germany}

\author{R. Cardoso}
\affiliation{Max-Planck-Institute for Chemical Physics of Solids, Dresden, Germany}

\author{F. Steglich}
\affiliation{Max-Planck-Institute for Chemical Physics of Solids, Dresden, Germany}

\author{M. Sigrist}
\affiliation{Institute for Theoretical Physics, ETH Zurich, Zurich, Switzerland}

\author{A. Prokofiev}
\affiliation {Institut f\"{u}r Festk\"{o}rperphysik, Technische Universit\"{a}t Wien, Wien, Austria}

\author{E. Bauer}
\affiliation {Institut f\"{u}r Festk\"{o}rperphysik, Technische Universit\"{a}t Wien, Wien, Austria}


\date{\today}

\begin{abstract}
We report on the vortex dynamics of a single crystal of the non-centrosymmetric heavy-fermion superconductor CePt$_{3}$Si. Decays of the remnant magnetization display a clean logarithmic time dependence with rates that follow the temperature dependence expected from the Kim-Anderson theory. The creep rates are lower than observed in any other centrosymmetric superconductor and are not caused by high critical currents. On the contrary, the critical current in  CePt$_{3}$Si is considerably lower than in other superconductors with strong vortex pinning indicating that an alternative impediment on the flux line motion might be at work in this superconductor.
\end{abstract}


\maketitle


Unconventional superconductors which violate spontaneously other symmetries beside the $ U(1)$-gauge symmetry have been found to show many intriguing properties. Among such superconductors Sr$_2$RuO$_4$, PrOs$_4$Sb$_{12}$ and possibly UPt$_3$ have been identified as time reversal symmetry breaking by means of zero field $\mu$SR studies \cite{Aoki}. These compounds show surprisingly slow vortex dynamics  with creep rates lower than in any other superconductor \cite{Amann, Dumont_Thesis, Cichorek}. It has been proposed that this behavior is connected with the presence of domain walls between different degenerate superconducting phases which would occur naturally in time reversal symmetry breaking states.
Such domain walls could act as barriers for vortices, rather than the pinning of vortices at impurities and defects  \cite{Sigrist}. The latter pinning mechanism would have implied very high critical currents unlike what was observed in the experiments.

Interestingly, our present investigation reveals extremely slow flux line dynamics without simultaneous large critical current also in CePt$_3$Si, a non-centrosymmetric heavy-fermion superconductor, discovered recently by Bauer et al \cite{Bauer}. This compound is a member of a whole class of presumably unconventional heavy-fermion superconductors such as CeRhSi$_3$ \cite{Kimura}, CeIrSi$_3$ \cite{Sugitani}, and UIr \cite{Akazawa} whose crystal lattices do not posses an inversion center. Among these systems,  CePt$_3$Si is the only one where superconductivity sets in already at ambient pressure. Superconductivity in systems without inversion symmetry has been discovered also outside the heavy-fermion class as for example in Li$_2$(Pd,Pt)$_3$B \cite{Togano,Badica} or Mg$_{10}$Ir$_{19}$B$_{16}$ \cite{Klimczuk}.


In  CePt$_3$Si antiferromagnetic order sets in at a N\'{e}el temperature $T_N=2.2$~K while the system adopts a superconducting ground state below a transition temperature $T_c=0.75$~K for polycrystalline samples \cite{Bauer}. Lower superconducting transition temperatures have been reported for single-crystals \cite{Takeuchi}. Long-range magnetic order coexists with superconductivity on a microscopic scale as revealed by $\mu$SR investigations \cite{Amato}. The upper critical field $H_{c2}\approx 3-5$~T exceeds the Pauli-Clogston limit $H_P\approx1.1$~T indicating that paramagnetic limiting is unimportant here. Knight shift data actually display no reduction of the spin susceptibility below $T_c$, for magnetic fields perpendicular or parallel to the crystallographic $c$-axis \cite{Yogi2006}. Power laws describing the low-temperature behavior of thermal conductivity \cite{Izawa}, penetration depth \cite{Bonalde}, $1/T_1$ relaxation rate \cite{Yogi2004}, and specific heat \cite{Takeuchi} observed in CePt$_3$Si suggests a superconducting gap with line nodes. Remarkably, CePt$_3$Si is the only heavy-fermion system to exhibit a Hebel-Slichter coherence peak below $T_c$  \cite{Yogi2004}, a feature characteristic to an s-wave superconductor.


In this letter we present an experimental investigation of flux dynamics on a single-crystal of CePt$_3$Si which reveals the presence of an unconventional and very effective vortex pinning mechanism.

The high-quality CePt$_3$Si single-crystal investigated was grown using a Bridgman technique and the sample was oriented, cut and polished in a parallelepiped  shape with the dimensions $4.60$~mm$\times 2.65$~mm$\times 1.05$~mm. The longer dimension is parallel to the crystallographic $a$-axis, while the smaller one is parallel to the $b$-axis. Prior to the flux creep measurements the sample was characterized by ac magnetic susceptibility and specific heat. The investigation of vortex dynamics was performed in a dilution refrigerator in the temperature range $0.1$~K$\leq T \leq 0.5$~K with the sample enclosed in a custom-built mixing chamber and using a SQUID detector to determine the magnetic flux expelled. The external magnetic field applied to drive the sample into the Bean critical state was applied along the $a$-axis.  In the same experimental configuration, ac susceptibility experiments were performed in the temperature range $0.025$~K$ \leq T \leq 2.4$~K using an inductance bridge with a SQUID as null detector. A very low ac excitation field of $H=1.3$~mOe was applied along the $a$-axis at a frequency $f=80$~Hz. The temperature dependence of the specific heat was measured in the temperature range $0.05$~K $\leq T \leq 4.5$~K employing a quasi-adiabatic pulse method.
\begin{figure}[t]
\centering
\includegraphics[angle=0,width=8.5cm,clip]{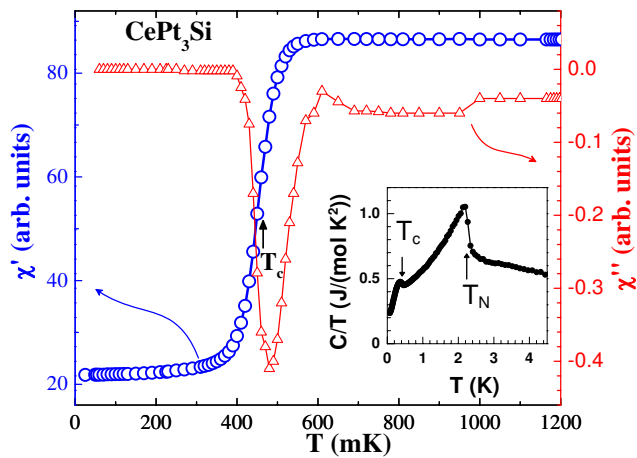}
\caption{\label{fig1} Temperature dependences of the real the imaginary part of the ac magnetic susceptibility across the superconducting phase transition. Inset: Temperature dependence of the specific heat divided by temperature.}
\end{figure}

Both, the real, $\chi'$, and the imaginary, $\chi''$, part of the ac-susceptibility (Fig. \ref{fig1}) clearly reveal the superconducting transition with the mid-point of the anomaly in $\chi'$ centered at $T_c=0.45$~K. The transition width defined as the temperature difference between the 10\% and 90\% drop of the real part of susceptibility across the transition $\Delta T = 0.1$~K is substantially smaller than the value observed for polycrystalline samples. However, our finding is in excellent agreement with previous studies on high quality single crystals \cite{Takeuchi}. The $T_c$ discrepancy between single- and polycrystals is not yet properly understood, but one possible explanation is that this compound has an homogeneity range \cite{Gribanov} similar for example to the well known case of CeCu$_2$Si$_2$ \cite{Steglich}, which allows for homogeneous samples with slightly different compositions but substantially different physical properties to form. Another scenario  \cite{Iniotakis} suggests that twin boundaries could enhance the trend to superconductivity in polycrystalline samples.
Upon warming up the sample in the normal state, no signature of the transition from the long range magnetically ordered state into the paramagnetic phase was detected in $ \chi' $ and $ \chi''$ up to $T=2.4$~K, for our field orientation ($H\parallel a$).
\begin{figure}[t]
\centering
\includegraphics[angle=0,width=8.5cm,clip]{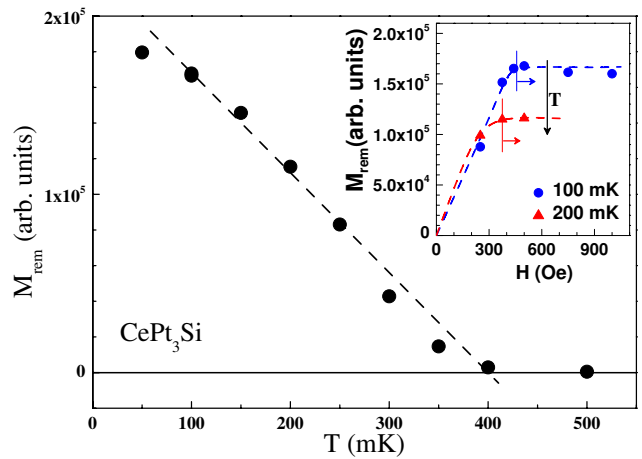}
\caption{\label{fig2} Temperature dependence of the total remnant magnetization. Dashed line is a linear fit to the data. Inset: Total remanent magnetization at $T=0.1$~K and $T=0.2$~K as function of the external magnetic field $H$.}
\end{figure}

The temperature dependence of the specific heat divided by temperature is depicted in the inset of Fig.\ref{fig1}.  The transition into the antiferromagnetically  ordered state is clearly visible as a sharp peak at $T_N=2.3$~K, a value consistent with the one obtained in previous specific heat studies \cite{Takeuchi}. Upon further cooling down, the system adopts a superconducting ground state at $T_c=0.42$~K, in good agreement with our susceptibility data and Ref. \cite{Takeuchi}. Both $T_N$ and $T_c$ are  defined as the mid point of the jump in $C$ across the respective anomaly. The $C/T$ data in the temperature range $0.5\leq T\leq 2.1$~K are well described by $C/T=423$~mJ/(mol K$^2)+140~T^2$~mJ/(mol K$^4$). We remove the phononic and antiferromagnetic contributions to the specific heat by subtracting $140~T^3$~mJ/(mol K$^4$) from the $C(T)$ data and obtain a normal state Sommerfeld coefficient $\gamma_n = 400$~mJ/(mol K$^2)$. This leads to a jump of the specific heat at the superconducting phase transition of $\Delta C / (\gamma_n T_c)\approx 0.29$, a value situated significantly below the BCS-theory prediction of $\Delta C /(\gamma_n T_c)=1.43$. In the superconducting state, $C$ exhibits a quadratic temperature dependence down to $T=0.1$~K rather than an exponential one, indicative of the existence of line nodes in the superconducting gap \cite{Samokhin}. A zero temperature interception of the quadratic specific heat would yield a residual electronic specific heat coefficient with a finite value of $\gamma_s = 145$~mJ/mol K$^2$. However, below $T=0.1$~K, the specific heat has a weaker temperature dependence, therefore the residual $\gamma_s$ will assume probably an even higher value.

Isothermal relaxation curves of the remanent magnetization $M_{rem}$ were taken after cycling the specimen in an external dc magnetic field $H$. Vortices were introduced into the sample at a constant and slow rate in order to avoid eddy current heating and using, at the lowest temperature, a magnetic field just high enough to drive the sample into the Bean critical state. The required magnetic field was kept constant in the sample for several minutes and than gradually reduced to zero.
\begin{figure}[t]
\centering
\includegraphics[angle=0,width=9cm,clip]{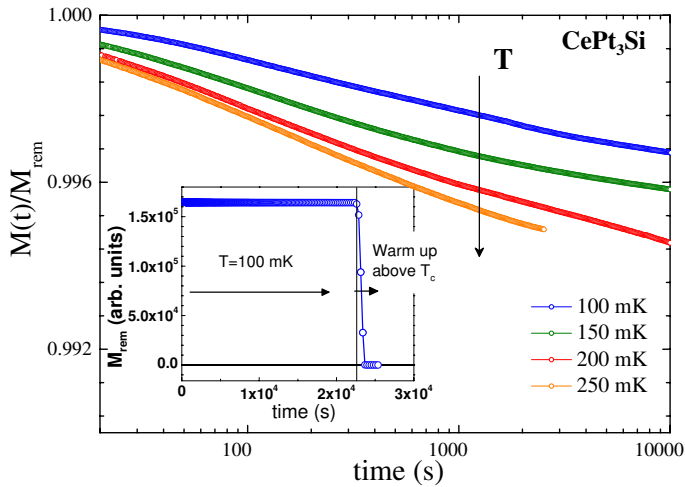}
\caption{\label{fig3} Normalized remanent magnetization as a function of time at different constant temperatures. Inset: remanent magnetization a as function of time at $T=0.1$~K. After $2.25\times 10^4$~s the sample is warmed up above $T_c$ and all the trapped magnetic flux is expelled.}
\end{figure}
Immediately after, the relaxation of the metastable magnetization was recorded with a digital flux counter for several hours. The time required to ramp down the field is negligible compared with the time of the relaxation measurement. The magnetic field was applied along the crystallographic $a$-direction. At the lowest temperature of our investigation,
$T = 100$~mK, we determined the field corresponding to the Bean critical state, $H_s$, as the field where the remanent magnetization saturates as a function of the applied external magnetic field. For this sample, we found $H_s=500$~Oe at $T=100$~mK (inset of Fig. \ref{fig2}). At higher temperatures the sample is in the critical state already for smaller external fields, since $H_s$ decreases upon increasing $T$ as demonstrated in the inset of Fig. \ref{fig2}. In the main part of Fig. \ref{fig2}, we present the temperature dependence of the remanent magnetization obtained after cycling the sample in a field of $H=500$~Oe. To obtain the value of $M_{rem}$, after cycling the field at constant temperature we warmed up the sample well above $T_c$ and recorded the magnetic flux expelled. $M_{rem}$ decreases monotonically upon increasing temperature with the experimental data well described by a linear fit  (dashed line in Fig. 2) which intercepts zero at around $T\approx 0.41$~K. This is in excellent agreement with the value of $T_c$ yielded by ac susceptibility and specific heat measurements. At $T=0.5$~K no flux was trapped in the crystal clearly showing that the bulk of the sample is well in the normal state at this temperature.

Isothermal decays of the remanent magnetization at different temperatures are depicted in Fig.~3.
At constant temperature the flux escaping the sample is recorded for typically more than $10^4$~s. Then the sample is heated up above $T_c$ so all the trapped field is expelled out of the sample (inset of Fig.~3). In this way we obtain the value of the total remanent magnetization as the sum of the amount of flux expelled in the first $10^4$~s plus the flux removed while crossing $T_c$. This value of $M_{rem}$ is then used to normalize the creep rate. At all temperatures the decays show  a clear logarithmic time dependence as predicted by the Kim-Anderson theory \cite{Anderson}.  The creep rate becomes faster upon increasing the temperature as expected for thermally activated flux motion.
\begin{figure}[t]
\centering
\includegraphics[angle=0,width=7cm,clip]{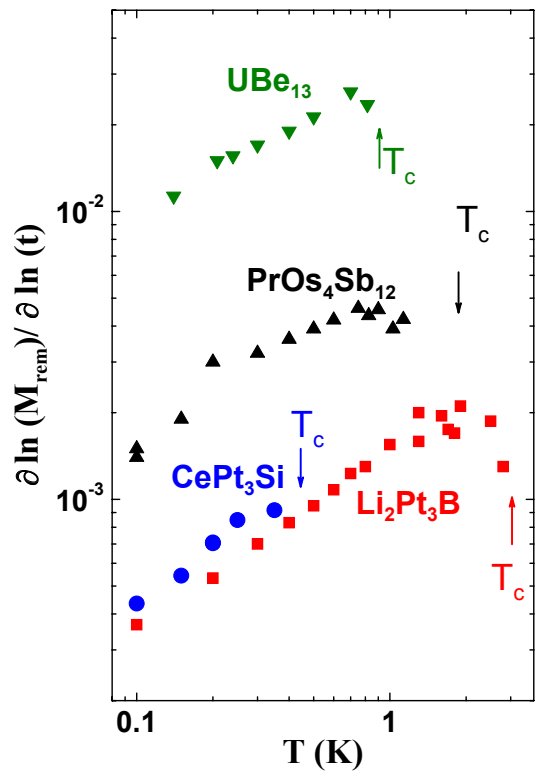}
\caption{\label{fig4}  Comparison of the normalized relaxation rates $S=\partial$ln$(M)/\partial$ln$(t)$  as function of temperature for different compounds in a log-log representation.}
\end{figure}
The temperature dependence of the normalized relaxation rates $S=\partial$ln$(M)/\partial$ln$(t)$ is depicted in Fig.~4 together with the rates obtained for the heavy-fermion superconductor UBe$_{13}$ \cite{Amann} which only brakes gauge symmetry, PrOs$_4$Sb$_{12}$ \cite{Cichorek} which also violates time reversal symmetry and the non-centrosymmetric superconductor Li$_2$Pt$_3$B  \cite{Miclea}. Remarkably, CePt$_3$Si has anomalously  small decay rates comparable only with  Li$_2$Pt$_3$B and lower by a factor of five than the very low creep rates observed in PrOs$_4$Sb$_{12}$. Li$_2$Pt$_3$B  breaks the inversion symmetry as well and displays extremely small creep rates. However, for the latter compound, in a certain temperature interval, the weak initial logarithmic creep is followed after several thousand seconds by a much faster, avalanche-like, also logarithmic, decay \cite{Miclea}. In general in superconductors with strong vortex pinning the critical current $j_c$ is high. However, this is not the case in CePt$_3$Si which has the lowest critical current among the compared superconductors (Fig.~5). The comparison depicted in Fig. 5 has been done for $T=300$~mK and in the framework of the Bean model which assumes a constant $j_c(T)\propto H_c(T)/d$, where $d$ is the thickness of the plate-like shaped single crystal.  A lower critical current for CePt$_3$Si is reflected in a reduced vortex density which could explain the lack of avalanche-like relaxation: vortices would need a time much longer than the experimental observation to exert a pressure high enough to overcome pinning barriers.

\begin{figure}[]
\centering
\includegraphics[angle=0,width=8.5cm,clip]{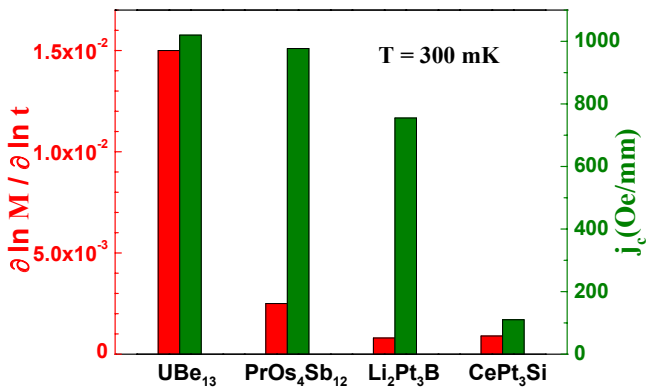}
\caption{\label{fig5}  Comparison of the  normalized relaxation rates $S=\partial$ln$(M)/\partial$ln$(t)$ and the critical current at $T=0.3$~K for different compounds.}
\end{figure}

The extremely slow vortex dynamics in CePt$_3$Si suggest an unconventional and very effective pinning mechanism. In contrast to UPt$_3$, Sr$_2$RuO$_4$, and PrOs$_4$Sb$_{12}$,
the superconducting phase of CePt$_3$Si conserves time reversal symmetry and the  intrinsic pinning mechanisms proposed for those \cite{Sigrist} do not apply here. On the other hand, in this non-centrosymmetric material crystalline twin boundaries separating twin domains of opposite non-centrosymmetricity could also provide the location for fractional vortices as suggested by Iniotakis et al.  \cite{Iniotakis}. Such twin boundaries would then also introduce a barrier for flux-line motion without affecting the critical current. A new refinement of the crystal structure of CePt$_3$Si, from X-ray intensity data collected on the same single crystal investigated in our study, shows a contribution of 87 \% of the main inversion twin component.

In conclusion, we observed extremely slow vortex dynamics in the non-centrosymmetric  CePt$_3$Si similar to Li$_2$Pt$_3$B. In both compounds the flux pinning is caused by an unconventional and very effective mechanism.  A possible explanation for this discovery, which might be  characteristic for a certain class of non-centrosymmetric superconductors could be the existence of fractionalized vortices on twin boundaries. However, this scenario needs independent verification apart from the flux dynamics reported here. No other explanations are known to us to date.
Unlike in Li$_2$Pt$_3$B we did not observed flux avalanches \cite{Miclea} which might be due to the much lower flux density reflected by the reduced critical current in  CePt$_3$Si.

We are grateful to the late Kazumi Maki for many enlightening and useful discussions on this subject. C.F.M would like to acknowledge the support of the German Research Foundation (DFG) under the auspices of the  MI~1171/1-1. ACM and MS have been financially supported by the Swiss Nationalfonds.  E.B is grateful to the Austrian FWF P18054.

\end{document}